\documentclass[a4paper,reqno]{amsart}

\usepackage{amsthm,amssymb,amsmath,amsfonts}
\usepackage{verbatim}
\usepackage{bbm}
\usepackage{hyperref}
\usepackage[all]{xy}
\usepackage{latexsym}
\usepackage{color}
\usepackage{tikz}
\usepackage[colorinlistoftodos]{todonotes} %Comments
\usepackage{graphicx}
\usepackage{caption}
\usepackage{subcaption}
\usepackage{tikz}
\usepackage{tikz-qtree}
\usepackage{verbatim}

\usepackage{rotating}

\usepackage{amsaddr}

\usepackage{natbib}

\newtheorem{theorem}{Theorem}

\newtheorem{corollary}[theorem]{Corollary}

\theoremstyle{definition}

\begin{document}

\title[Moments and Entropy of the Interpolating Family]{Moments and Entropy of the Interpolating Family of Size Distributions}

\author[Corinne Sinner and Patrick Weber]{Corinne Sinner and Patrick Weber}
\address{Department of Mathematics, Universit\'e libre de Bruxelles (ULB), Belgium. E-Mail: csinner@ulb.ac.be, pweber@ulb.ac.be.}

\maketitle

%\vspace{-1cm}

\begin{abstract}
\cite{interpolating_family} recently introduced a five-parameter family of size distributions, coined \textsl{Interpolating Family} or \textsl{IF~distribution} for short. In this complementary note, we take advantage of the tractability of the IF~distribution to compute the moments and the differential entropy. As a consequence, we deduce at a single stroke the corresponding expressions for many well-known size distributions arising as special cases of the IF distribution.
\end{abstract}

\hfill

\noindent
{\it Keywords:} Interpolating Family, Moments, Differential Entropy.

%%%%%%%%%%%%%%%%%%%%%%%%%%%%%%%%%%%%%%%%%%%%%%%%%%%%%%%%%%%%%%%%%%%

\section{Introduction}
\label{sec:Intro}
%In Sinner et al.~(2016), the authors recently introduced a five-parameter family of size distributions, coined \textsl{interpolating family} or \textsl{IF~distribution} for short. In this complementary short note, we take advantage of the tractability of the IF distribution to compute the moments and the differential entropy. As a consequence, we deduce at a single stroke the corresponding expressions for many well-known size distributions arising as special cases of the IF distribution.

%\section{The interpolating family}
%\label{}
\cite{interpolating_family} recently introduced a five-parameter family of size distributions, called \textsl{Interpolating Family} or \textsl{IF distribution}. In this note, we give explicit expressions for  the moments and the differential entropy of its three main subfamilies, denoted IF1, IF2 and IF3 distributions. %, derived its stochastic and inferential properties and provided three applications to real data sets.
%
%\hfill
%
%nach intro

\hfill

The probability density function (pdf) of the IF distribution is given by
\begin{equation*}
	f_{p,b}(x)= \text{sign}(b)q~g(x) G(x)^{-q-1} \left(1-\frac{1}{p+1}G(x)^{-q}\right)^p, 
\end{equation*}
where $G(x)=(p+1)^{-\frac{1}{q}}+\left(\frac{x-x_0}{c}\right)^b$, $g(x)=\frac{\mathrm{d}}{\mathrm{d}x}G(x)=\frac{b}{c}\left(\frac{x-x_0}{c}\right)^{b-1}$ and $x \geqslant x_0$. The distribution depends on five parameters: an interpolation parameter $p \in [0,\infty]$, a shape parameter $b \in \mathbb{R}_0$, a scale parameter $c>0$, a tail-weight parameter $q>0$ and a location parameter $x_0 \geqslant 0$.
Three interesting four-parameter subfamilies are nested inside this five-parameter family. If $p=0$, we get a power law distribution, called \textit{Interpolating Family of the first kind (IF1)} with pdf
\begin{equation*}
	f_{0,b}(x)=\frac{|b|q}{c} \left(\frac{x-x_0}{c}\right)^{b-1} \left(1+\left(\frac{x-x_0}{c}\right)^{b}\right)^{-q-1},
\end{equation*}
where $x\geqslant x_0$. The IF1 distribution contains many well-known distributions such as for example the Pareto type I, II, III and IV, Lomax, Fisk and Burr type III and~XII distributions~\citep{Burr42, Tadikamalla, Lindsay}. If, on the other hand, $p \to \infty$, we get a power law distribution with exponential cut-off, called \textit{Interpolating Family of the second kind (IF2)} with pdf
\begin{equation*}
	f_{\infty,b}(x)= \frac{|b|q}{c} \left(\frac{x-x_0}{c}\right)^{-bq-1} e^{-\left(\frac{x-x_0}{c}\right)^{-bq}}, 
\end{equation*}
where $x \geqslant x_0$. Special cases of the IF2 distribution are the Weibull, Fr\'{e}chet, Gumbel type II, Rayleigh and Exponential distributions.
Finally, if $b=1$, we get the \textit{Interpolating Family of the third kind (IF3)} with pdf
\begin{equation*}
	f_{p,1}(x)=\frac{q}{c} \left( (p+1)^{-\frac{1}{q}}+ \frac{x-x_0}{c}\right)^{-q-1}  \left(1-\left( 1+ (p+1)^{\frac{1}{q}} \frac{x-x_0}{c}\right)^{-q} \right)^p, 
\end{equation*}
where $x \geqslant x_0$. The IF3 distribution contains most notably the Generalized Lomax and the Stoppa distributions \citep{KleiberKotz}. 
%The aim of this note is to compute the moments of these three subfamilies IF1, IF2 and IF3 as well as the entropy of the IF distribution.

%\hfill
%
%The purpose of this note is twofold. Firstly, we compute in Section~\ref{sec:Moments} the moments of the three subfamilies IF1, IF2 and IF3 and summarize in Section~\ref{sec:MeanSpecialCases} the means for various well-known special cases of the IF distribution. Secondly, we compute in Section~\ref{sec:Entropy} the differential entropy of the IF distribution and assemble the corresponding entropies of the special cases in Section~\ref{sec:EntropySpecialCases}.

%%%%%%%%%%%%%%%%%%%%%%%%%%%%%%%%%%%%%%%%%%%%%%%%%

\section{Moments}
\label{sec:Moments}

%%%%%%%%%%%%%%%%%%%%%%%%%%%%%%%%%%%%%%%%%%%%%%%%%

The $r^\text{th}$ moment of the IF distribution is given by 
\begin{equation*}
	\begin{array}{ccl}
		\mathbb{E}\left[X^r\right]&=& \int \limits_{x_0}^\infty x^r~f_{p,b}(x)~\mathrm{d}x.
	\end{array}
\end{equation*}
If we make the change of variables $y= (p+1)^{-\frac{1}{q}} + \left(\frac{x-x_0}{c}\right)^b$ and apply Newton's binomial theorem, then we get
\begin{equation*}
	\begin{array}{ccl}
		\mathbb{E}\left[X^r\right]&=& \sum \limits_{i=0}^r {r \choose i} x_0^i c^{r-i} \underbrace{ \int \limits_{(p+1)^{-\frac{1}{q}}}^\infty q~y^{-q-1} \left( y-(p+1)^{-\frac{1}{q}} \right)^{\frac{r-i}{b}} \left( 1 - \frac{y^{-q}}{p+1} \right)^p \mathrm{d}y.}_{I(p,b,q)}
	\end{array}
\end{equation*}
We compute the integral $I(p,b,q)$ for the three subfamilies IF1, IF2 and~IF3.

\subsection*{IF1 distribution}
If we plug in $p=0$ and set $z=\frac{1}{y}$, then we get
\begin{equation*}
	\begin{array}{ccl}
		 I(0,b,q) &=& \int \limits_1^\infty q~y^{-q-1} \left( y-1 \right)^{\frac{r-i}{b}} \mathrm{d}y = q \int \limits_0^1 z^{q-1-\frac{r-i}{b}} \left(1-z\right)^{\frac{r-i}{b}} \mathrm{d}z.
	\end{array}
\end{equation*}	
If either $b>0$ and $r<bq$ or else $b<0$ and $r<-b$, then this can be written as
\begin{equation*}
	\begin{array}{ccl}
		I(0,b,q) &=& q B(q-\frac{r-i}{b},1+ \frac{r-i}{b}) = \frac{\Gamma\left(q-\frac{r-i}{b}\right) \Gamma\left(1+ \frac{r-i}{b}\right)}{\Gamma(q)}.
	\end{array}
\end{equation*}
The $r^\text{th}$ moment of the IF1 distribution is given by
\begin{equation*}
	\begin{array}{ccl}
		\mathbb{E}\left[X^r\right]&=& \sum \limits_{i=0}^r {r \choose i} x_0^i c^{r-i} \frac{\Gamma\left(q-\frac{r-i}{b}\right) \Gamma\left(1+ \frac{r-i}{b}\right)}{\Gamma(q)} \quad \text{if } \left\{
\begin{array}{l}
b>0 \text{ and } r<bq,\\
b<0 \text{ and } r<-b.
\end{array}
\right.
	\end{array}
\end{equation*}

\subsection*{IF2 distribution}
If we make the change of variables $z=y^{-q}$, then we get
\begin{equation*}
	\begin{array}{ccl}
		\lim \limits_{p \to \infty} I(p,b,q) &=& \lim \limits_{p\to \infty} \int \limits_0^{p+1} \left(z^{-\frac{1}{q}} - (p+1)^{-\frac{1}{q}} \right)^{\frac{r-i}{b}} \left( 1- \frac{z}{p+1} \right)^p \mathrm{d}z.
	\end{array}
\end{equation*}
We may apply the Lebesgue dominated convergence theorem to deduce that
\begin{equation*}
	\begin{array}{ccl}
		\lim \limits_{p \to \infty} I(p,b,q) &=& \int \limits_0^\infty z^{-\frac{r-i}{bq}} e^{-z}~\mathrm{d}z = \Gamma\left(1- \frac{r-i}{bq}\right).
	\end{array}
\end{equation*}
The $r^\text{th}$ moment of the IF2 distribution is given by
\begin{equation*}
	\begin{array}{ccl}
		\mathbb{E}\left[X^r\right]&=& \sum \limits_{i=0}^r {r \choose i} x_0^i c^{r-i}~\Gamma\left(1- \frac{r-i}{bq}\right) \quad \text{if } \left\{
		\begin{array}{l}
b>0 \text{ and } r<bq,\\
b<0. 
\end{array}
\right.
	\end{array}
\end{equation*}

\subsection*{IF3 distribution}
If $b=1$, then we make the change of variables $z=\frac{y^{-q}}{p+1}$. Using Newton's binomial theorem, we get
\begin{equation*}
	\begin{array}{ccl}
		I(p,1,q) &=& (p+1)^{1-\frac{r-i}{q}} \int \limits_0^1 \left(z^{-\frac{1}{q}}-1\right)^{r-i} \left(1-z\right)^p \mathrm{d}z \\
		&=& (p+1)^{1-\frac{r-i}{q}} \sum \limits_{k=0}^{r-i} {r-i \choose k} (-1)^k \int \limits_0^1 z^{-\frac{1}{q}(r-i-k)}\left(1-z\right)^p\mathrm{d}z.
	\end{array}
\end{equation*}
If $r<q$, then this can be written as
\begin{equation*}
	\begin{array}{ccl}
		I(p,1,q) &=& (p+1)^{1-\frac{r-i}{q}} \sum \limits_{k=0}^{r-i} {r-i \choose k} (-1)^k B\left(1-\frac{1}{q}(r-i-k),p+1\right).	\end{array}
\end{equation*}
Under the hypothesis $r<q$, the $r^\text{th}$ moment of the IF3 distribution is given by
\begin{equation*}
	\begin{array}{ccl}
		\mathbb{E}\left[X^r\right]&=& \sum \limits_{i=0}^r {r \choose i} x_0^i c^{r-i}~(p+1)^{1-\frac{r-i}{q}} \sum \limits_{k=0}^{r-i} {r-i \choose k} (-1)^k B\left(1-\frac{1}{q}(r-i-k),p+1\right).
	\end{array}
\end{equation*}

%%%%%%%%%%%%%%%%%%%%%%%%%%%%%%%%%%%%%%%%%%%%%%%%%%%%%

\section{Differential entropy}
\label{sec:Entropy}

The differential entropy $h$ of a continuous random variable with pdf $f$ is %defined as
\begin{equation}
	h(f) = - \int\limits_S f(x) \ln(f(x)) \textrm{d}x, \label{Entropy}
\end{equation}
where $S$ is the support of the random variable \citep{ShannonEntropy}. 
We will show that the differential entropy of the IF distribution is given by
\begin{equation*}
h(f_{p,b}) =-\ln\left(\frac{|b|q}{c}\right) -\frac{b-1}{b} F(p,q) - \frac{bq+1}{bq} \ln(p+1)  + \frac{q+1}{q} \text{H}_{p+1}+ \frac{p}{p+1},
\end{equation*}
where $H_{p+1} = \sum \limits_{k=1}^{p+1} \frac{1}{k}$ is the $(p+1)^{\text{th}}$ harmonic number and 
\begin{equation*}
F(p,q) = (p+1) \int \limits_0^1 \ln \left(t^{-\frac{1}{q}} -1\right) \left(1-t\right)^p \mathrm{d}t.
\end{equation*}
To prove this result, we proceed as follows. 
%Firstly, as the differential entropy is translation invariant, we may fix $x_0=0$. Secondly, using the scale relation $h\left(f_{p,b}(x;c,q,x_0)\right)= h\left(f_{p,b}(x;1,q,x_0)\right) + \ln(c)$ we may also set $c=1$.
If ${0\leqslant p<\infty}$, then we apply the change of variables $t=\left(1+(p+1)^{\frac{1}{q}}\left(\frac{x-x_0}{c}\right)^b\right)^{-q}$ to~\eqref{Entropy}:
\begin{equation*}
\begin{array}{ccl}
	h(f_{p,b})&=&-\ln\left(\frac{|b|q}{c}\right)-\frac{b-1}{b} (p+1)\int \limits_{0}^{1} \ln(t^{-\frac{1}{q}}-1) (1-t)^p~\mathrm{d}t \\
			&& -\frac{bq+1}{bq} \ln(p+1) (p+1) \int \limits_{0}^{1} (1-t)^p~\mathrm{d}t  \\
			&& - \frac{q+1}{q} (p+1) \int \limits_{0}^{1} \ln(t) (1-t)^p \mathrm{d}t  -p (p+1) \int \limits_{0}^{1} \ln(1-t) (1-t)^p~\mathrm{d}t.
\end{array}			
\end{equation*}
Clearly, $(p+1) \int \limits_0^1 (1-t)^p \mathrm{d}t =1$ and $(p+1) \int \limits_0^1 \ln(1-t) (1-t)^p \mathrm{d}t=-\frac{1}{p+1}$. Moreover, integrating by parts, we deduce that $(p+1) \int \limits_0^1 \ln(t) (1-t)^p \mathrm{d}t = -H_{p+1}$ and the result follows.
%As a consequence, by setting $p=0$, the differential entropy of the IF1 distribution is given by
In particular, the differential entropy of the IF1 distribution~($p=0$) is given by
\begin{eqnarray*}
h(f_{0,b}) = -\ln\left(\frac{|b|q}{c}\right) + \frac{b-1}{b}H_{q-1} + \frac{q+1}{q}
\end{eqnarray*}
and the differential entropy of the IF3 distribution~($0<p<\infty$ and $b=1$) is
\begin{equation*}
h(f_{p,1})= - \ln\left(\frac{q}{c}\right) + \frac{q+1}{q}\left(H_{p+1} - \ln (p+1)\right) + \frac{p}{p+1}.
\end{equation*}
On the other hand, if $p \to \infty$, then we make the change of variables $z=\left(\frac{x-x_0}{c}\right)^{-bq}$ in~\eqref{Entropy} and get
\begin{equation*}
h(f_{\infty,b}) = - \ln\left(\frac{|b|q}{c}\right) -\frac{bq+1}{bq} \int_{0}^\infty  e^{-z} \ln(z)~\mathrm{d}z + \int_{0}^\infty z e^{-z}~\mathrm{d}z.
\end{equation*}
We can express the integral in the second term as the Euler--Mascheroni constant
\begin{equation*}
\int_0^\infty  e^{-z} \ln(z)~\mathrm{d}z = - \gamma_E
\end{equation*}
and the integral in the third term can be simplified to
\begin{equation*}
\int_0^\infty z e^{-z}~\mathrm{d}z = \Gamma(2) =1.
\end{equation*}
We conclude that the differential entropy of the IF2 distribution ($p\to \infty$) becomes
\begin{equation*}
h(f_{\infty,b}) = - \ln\left(\frac{|b|q}{c}\right) + \frac{bq+1}{bq} \gamma_E + 1,
\end{equation*}
where $\gamma_E = \lim \limits_{p \to \infty} \left( H_{p+1} - \ln(p+1) \right)$.

\hfill

By convexity~\citep{handbook_entropy}, we deduce that the IF subfamilies maximize the differential entropy within the class of all continuous probability distributions under the following constraints:
\begin{corollary}
The IF1 distribution maximizes the differential entropy within the class of all continuous probability distributions satisfying the constraints
\begin{equation*}
\begin{array}{lcl}
\mathbb{E}\left[\ln(\frac{x-x_0}{c})\right] &=& - \frac{H_{q-1}}{b} \\
\mathbb{E}\left[\ln(1+(\frac{x-x_0}{c})^b) \right] &=& \frac{1}{q}
\end{array}
\end{equation*}
and having support $x\geqslant x_0$.
\end{corollary}

\begin{corollary}
The IF2 distribution maximizes the differential entropy within the class of all continuous probability distributions satisfying the constraints
\begin{equation*}
\begin{array}{lcl}
\mathbb{E}\left[\ln(\frac{x-x_0}{c})\right] &=&\frac{\gamma_E}{bq} \\
\mathbb{E}\left[\left(\frac{x-x_0}{c}\right)^{-bq}\right] &=& 1
\end{array}
\end{equation*}
and having support $x \geqslant x_0$.
\end{corollary}

\begin{corollary}
The IF3 distribution maximizes the differential entropy within the class of all continuous probability distributions satisfying the constraints
\begin{equation*}
\begin{array}{lcl}
\mathbb{E}\left[\ln\left( (p+1)^{-\frac{1}{q}} + \frac{x-x_0}{c} \right) \right] &=& \frac{1}{q} \left( H_{p+1} - \ln(p+1) \right)\\
\mathbb{E}\left[ \ln \left( 1- \left( 1 + (p+1)^{\frac{1}{q}}\frac{x-x_0}{c} \right)^{-q} \right)      \right] &=& - \frac{1}{p+1} 
\end{array}
\end{equation*}
and having support $x \geqslant x_0$.
\end{corollary}

\section{Mean and entropy of special cases}

In the two following Tables, we display the mean and the differential entropy for some of the size distributions arising as special cases of the Interpolating Family. While most of these already appeared somewhere in the literature (see for example \cite{handbook_entropy} and~\cite{pareto_entropy}), we find it instructive to assemble them as done in the following Tables.\\

%While these maximum entropy constraints are not really useful in practice, they may nevertheless serve to find interesting unknown special cases within the interpolating family. If, for example, we set $b=q=1$, the maximum entropy constraints simplify to:
%\begin{equation*}
%\begin{array}{lcl}
%\mathbb{E} \left[\ln(\frac{x-x_0}{c} + \frac{1}{p+1}) \right] &=& 1 \\
%\mathbb{E} \left[\ln(\frac{x-x_0}{c}) \right] &=& \frac{p}{p+1}.
%\end{array}
%\end{equation*} 

\newpage

%\section{Mean of special cases}
%\label{sec:MeanSpecialCases}
%
%In this Section we give the mean for some of the size distributions arising as special cases of the Interpolating Family. While all of these are well-known, we find it instructive to assemble them as done in the following Table.\\

\begin{center}
\begin{sideways}
	\begin{tabular}{|c||c||c|c|c|c|}
	\hline
	Distribution & \# & Parameters & Mean & Constraint \\
	name & par. & $(p,b,c,q,x_0)$ & $\mathbb{E}\left[X\right]$ & \\
	\hline
	\hline
	Pareto IV & 4 &$(0,\frac{1}{\gamma}>0,c,q,x_0)$ & $x_0+c\frac{\Gamma\left(q-\gamma \right)\Gamma\left(1+\gamma \right)}{\Gamma(q)}$& $q>\gamma$ \\
	\hline
	Lindsay--Burr III & 4 &$(0,b<0,c,q,x_0)$ & $x_0+c\frac{\Gamma\left(q-\frac{1}{b}\right)\Gamma\left(1+\frac{1}{b}\right)}{\Gamma(q)}$& $b<-1$ \\
	\hline
	Pareto II & 3 & $(0,1,c,q,x_0)$ & $x_0+\frac{c}{q-1}$ & $q>1$\\
	\hline
	Pareto III & 3 & $(0,\frac{1}{\gamma}>0,c,1,x_0)$  & $x_0 + c \Gamma\left(1-\gamma \right)\Gamma\left(1+\gamma \right)$& $\gamma<1$\\
	\hline 
	Tadikamalla--Burr XII & 3 & $(0,b>0,c,q,0)$ & $c \frac{\Gamma\left(q-\frac{1}{b}\right)\Gamma\left(1+\frac{1}{b}\right)}{\Gamma(q)} $ & $bq>1$  \\
	\hline
	Fisk & 2 & $(0,b>0,c,1,0)$ &  $c \Gamma\left(1-\frac{1}{b}\right) \Gamma\left(1+\frac{1}{b}\right)$ &$b>1$ \\
	\hline
	Lomax & 2 & $(0,1,c,q,0)$ & $\frac{c}{q-1}$ & $q>1$\\
	\hline
	Pareto I & 2 & $(0,1,x_0,q,x_0)$ & $\frac{q}{q-1}x_0$& $q>1$\\
	\hline
	Burr XII  & 2 & $(0,b>0,1,q,0)$& $\frac{\Gamma\left(q-\frac{1}{b}\right)\Gamma\left(1+\frac{1}{b}\right)}{\Gamma(q)} $& $bq>1$  \\
	\hline
	\hline
	%IF2 & 4 & $(\infty, b,c,q,x_0)$ & $x_0+ c\Gamma\left(1-\frac{1}{bq}\right)$ & $bq>0$ \\
	%\hline
	Weibull & 3 & $(\infty,-1,c,q,x_0)$&$x_0+c\Gamma\left(1+\frac{1}{q}\right)$ & \\
	\hline
	Fr\'{e}chet & 3 & $(\infty,1,c,q,x_0)$ & $x_0+ c\Gamma\left(1-\frac{1}{q}\right)$ & $q>1$\\
	\hline
	Gumbel II & 2 & $(\infty,1,c,q,0)$ & $c\Gamma\left(1-\frac{1}{q}\right)$ & $q>1$\\
	\hline
	Rayleigh & 1 &$(\infty,-1,c,2,0)$ & $\frac{c}{\sqrt{2}} \sqrt{\frac{\pi}{2}}$ & \\
	\hline
	Exponential & 1 & $(\infty,-1,c,1,0)$& $c$ & \\
	\hline
	\hline
	%IF3 & 4 & $(m-1,1,c,q,x_0)$ & & \\
	%\hline
	Generalized Lomax & 3 &$(m-1,1,c,q,0)$& $c~m^{1-\frac{1}{q}} \left(B\left(1-\frac{1}{q},m\right)-B\left(1+\frac{1}{q},m\right)\right)$ &$q>1$ \\
	\hline
	Stoppa & 3 & $(m-1,1,c,q,cm^{-\frac{1}{q}})$ & $x_0~m~B\left(1-\frac{1}{q},m\right)$&$q>1$ \\
	\hline
	\end{tabular}
\end{sideways}
\end{center}

%\vspace{-0.8em}
%
%\section{Differential entropy of special cases}
%\label{sec:EntropySpecialCases}
%
%Similarly we display the differential entropy and the maximum entropy constraints for the special cases of the Interpolating Family. Most of these already appeared in the literature and we refer the reader to \cite{handbook_entropy} and~\cite{pareto_entropy}.

\begin{center}
\begin{sideways}
	\begin{tabular}{|c||c|c|c|c|}
	\hline
	Distribution & Parameters & Entropy & \multicolumn{2}{|c|}{Maximum entropy} \\
	name & $(p,b,c,q,x_0)$ & $h(f)$ & \multicolumn{2}{|c|}{constraints} \\
	\hline
	\hline
	Pareto IV &$(0,\frac{1}{\gamma}>0,c,q,x_0)$ & $(1-\gamma)H_{q-1}+\frac{q+1}{q}-\ln \left(\frac{q}{c\gamma}\right)$ & $\mathbb{E}[\ln(\frac{x-x_0}{c})]=-\gamma H_{q-1}$& $\mathbb{E}[\ln(1+(\frac{x-x_0}{c})^{\frac{1}{\gamma}})]=\frac{1}{q}$\\
	\hline
	Lindsay--Burr III &$(0,b<0,c,q,x_0)$ & $\frac{b-1}{b} H_{q-1}+\frac{q+1}{q}-\ln \left(\frac{|b|q}{c}\right)$&$\mathbb{E}[\ln(\frac{x-x_0}{c})]=-\frac{H_{q-1}}{b}$ &$\mathbb{E}[\ln(1+(\frac{x-x_0}{c})^b)]=\frac{1}{q}$ \\
	\hline
	Pareto II & $(0,1,c,q,x_0)$ &$\frac{q+1}{q}-\ln\left(\frac{q}{c}\right)$ & $\mathbb{E}[\ln(\frac{x-x_0}{c})]=-H_{q-1}$&$\mathbb{E}[\ln(1+\frac{x-x_0}{c})]=\frac{1}{q}$\\
	\hline
	Pareto III & $(0,\frac{1}{\gamma}>0,c,1,x_0)$  &$2+\ln(c \gamma)$ &$\mathbb{E}[\ln(\frac{x-x_0}{c})]=0$ &$\mathbb{E}[\ln(1+(\frac{x-x_0}{c})^{\frac{1}{\gamma}})]=1$ \\
	\hline 
	Tadikamalla--Burr XII & $(0,b>0,c,q,0)$ & $\frac{b-1}{b}H_{q-1}+\frac{q+1}{q}-\ln(\frac{|b|q}{c})$&$\mathbb{E}[\ln(\frac{x}{c})]= -\frac{H_{q-1}}{b}$& $\mathbb{E}[\ln(1+(\frac{x}{c})^b)]=\frac{1}{q}$\\
	\hline
	Fisk & $(0,b>0,c,1,0)$ & $2-\ln(\frac{b}{c})$& $\mathbb{E}[\ln(\frac{x}{c})]=0$&$\mathbb{E}[\ln(1+(\frac{x}{c})^b)]=1$ \\
	\hline
	Lomax & $(0,1,c,q,0)$ &$\frac{q+1}{q}-\ln(\frac{q}{c})$ &$\mathbb{E}[\ln(\frac{x}{c})]=-H_{q-1}$ &$\mathbb{E}[\ln(1+\frac{x}{c})]=\frac{1}{q}$\\
	\hline
	Pareto I & $(0,1,x_0,q,x_0)$ &$\frac{q+1}{q}-\ln(\frac{q}{x_0})$ &$\mathbb{E}[\ln(\frac{x}{c}-1)]=-H_{q-1}$ &$\mathbb{E}[\ln(\frac{x}{x_0})]=\frac{1}{q}$\\
	\hline
	Burr XII & $(0,b>0,1,q,0)$&$\frac{b-1}{b}H_{q-1}+\frac{q+1}{q}-\ln(bq)$ &$\mathbb{E}[\ln(x)]=-\frac{H_{q-1}}{b}$ & $\mathbb{E}[\ln(1+x^b)]=\frac{1}{q}$ \\
	\hline
	\hline
	%IF2 & 4 & $(\infty, b,c,q,x_0)$ & $x_0+ c\Gamma\left(1-\frac{1}{bq}\right)$ & $bq>0$ \\
	%\hline
	Weibull & $(\infty,-1,c,q,x_0)$& $\frac{q-1}{q}\gamma_{E}+1-\ln(\frac{q}{c})$&$\mathbb{E}[\ln(\frac{x-x_0}{c})]=-\frac{\gamma_E}{q}$ &$\mathbb{E}[(x-x_0)^q]=c^q$ \\
	\hline
	Fr\'{e}chet & $(\infty,1,c,q,x_0)$ &  $\frac{q+1}{q}\gamma_{E}+1-\ln(\frac{q}{c})$& $\mathbb{E}[\ln(\frac{x-x_0}{c})]=\frac{\gamma_E}{q}$ &$\mathbb{E}[(x-x_0)^{-q}]=c^{-q}$\\
	\hline
	Gumbel II & $(\infty,1,c,q,0)$ & $\frac{q+1}{q}\gamma_{E}+1-\ln(\frac{q}{c})$ & $\mathbb{E}[\ln(\frac{x}{c})]=\frac{\gamma_E}{q}$& $\mathbb{E}[x^{-q}]=c^{-q}$\\
	\hline
	Rayleigh &$(\infty,-1,c,2,0)$ &$\frac{1}{2}\gamma_E+\ln(\frac{c}{2})+1$ &$\mathbb{E}[\ln(\frac{x}{c})]=-\frac{\gamma_E}{2}$ &$\mathbb{E}[x^{2}]=c^2$ \\
	\hline
	Exponential & $(\infty,-1,c,1,0)$&$\ln(c)+1$ & &$\mathbb{E}[x]=c$ \\
	\hline
	\hline
	%IF3 & 4 & $(m-1,1,c,q,x_0)$ & & \\
	%\hline
	Generalized Lomax &$(m-1,1,c,q,0)$&$\frac{q+1}{q}(H_m-\ln(m))+\frac{m-1}{m}-\ln(\frac{q}{c})$ & $\mathbb{E}[\ln(m^{-\frac{1}{q}}+\frac{x}{c})]=\frac{1}{q}(H_m-\ln(m))$&$\mathbb{E}[\ln(1-(1+m^{\frac{1}{q}}\frac{x}{c})^{-q})]=-\frac{1}{m}$ \\
	\hline
	Stoppa& $(m-1,1,c,q,cm^{-\frac{1}{q}})$ &$\frac{q+1}{q}(H_m-\ln(m))+\frac{m-1}{m}-\ln(\frac{q}{c})$ & $\mathbb{E}[\ln(\frac{x}{c})]=\frac{1}{q}(H_m-\ln(m))$&$\mathbb{E}[\ln(1-(m^{-1}\frac{x}{c})^{-q})]=-\frac{1}{m}$ \\
	\hline
	\end{tabular}
\end{sideways}
\end{center}

\section{Conclusion}
The first aim of this complementary note was to provide closed form expressions for the moments of the IF1, IF2 and IF3 distributions introduced in \cite{interpolating_family}. The second aim was to calculate the differential entropy of the IF~distribution and of its subfamilies.

\bibliographystyle{apalike}
\bibliography{entropyrefs}

\end{document}